\documentclass{article}

\usepackage{graphicx}
\usepackage{psfig}
\usepackage{epsfig}
\usepackage[round]{natbib}

\title{Making use of the information in ensemble weather forecasts:
comparing the end to end and full
statistical modelling approaches}

\author{Stephen Jewson\footnote{\emph{Correspondence address}: RMS, 10 Eastcheap,
London, EC3M 1AJ, UK. Email: \texttt{x@stephenjewson.com}}}

\begin{document}

\maketitle

\begin{abstract}
We discuss how ensemble weather forecasts can be used, and
highlight the advantages and disadvantages of two particular
methods.
\end{abstract}

\section{Introduction}

Traditionally weather forecasts have been generated from single integrations
of weather simulation models.
Recently, however, ensemble integrations have also been used to generate
weather forecasts.
Ensemble integrations are created from the same weather simulation
models as are run for the single integrations but run in parallel in a number of different configurations. Each
configuration gives a slightly different forecast. From the point
of view of forecasting temperature, the main reason for the ensemble
approach to weather forecasting is that the average of the
forecasts from the different models is a better prediction than
the forecasts from the individual models themselves.

There are also other possible benefits from ensemble forecasts such as:
\begin{enumerate}

    \item that the spread across the ensemble could be used to give
    a useful indication of the uncertainty in the forecast
    \footnote{This is often assumed to be true, although as
    appropriately sceptical empiricists we observe that this has never
    been proven. In other words forecasts derived using the spread and past forecast error
    statistics have never been shown to
    be better than forecasts derived using past forecast error statistics alone.
    Our own research (\citet{jewson03g} and \citet{jewsondh03a})
    has so far failed to show any material benefit of using
    the spread as a predictor in either weather or seasonal forecasts}

    \item that the ensemble spread could be used to predict the width
    of the distribution of future forecast changes~\citep{jewsonz02a}

    \item that the temporal correlations in the ensemble could be used
    to predict temporal correlations~\citep{jewson03aa}

    \item that the spatial correlations in the ensemble could be used
    to predict spatial correlations

\end{enumerate}

Given all this potential information,
how should such ensemble weather forecasts be used by entities who
wish to predict the distribution of some variable that depends on the weather?
We will discuss the pros and cons
of two methods: the first is the
end to end approach (\citet{palmer02}), and the second we will call the
full statistical modelling approach.

\section{End to end use of ensemble forecasts}

\subsection{Uncalibrated end to end use of ensemble forecasts}

Uncalibrated end to end use of ensemble forecasts works as follows.
Each ensemble member from an ensemble forecast is converted
directly into a variable of user interest.
This gives a number of values for the user variable, and
these are then interpreted as samples from an estimate of
the future distribution of this variable.

This approach is not accurate, as is well known, because
the weather simulations in the ensemble need extensive calibration
before they can be considered as good estimates of the future
weather. This leads to the following method.

\subsection{Calibrated end to end use of ensemble forecasts}

The problem of lack of calibration in the ensemble members can
be partly overcome. If we consider temperature, which
is reasonably close to normally distributed, then the mean and the
standard deviation of the ensemble can be adjusted rather easily
(using calibration methods such as those described in~\citet{jewson04l})
while still preserving the individual ensemble members.
The calibrated ensemble members can then be converted into the
variable of user interest.

This method would be appropriate in some cases. However, it
has two limitations. These are discussed below.

\subsubsection{The correlation problem}
For some users of weather forecasts the temporal correlation structure
of the weather is important (in addition to the marginal distribution
of weather at a fixed time). As an example, consider a business that
cares about the distribution of possible amounts of money that it will
lose due to the weather over the next ten days, and imagine that
it is the number of freezing days that drives loss.
The autocorrelation in weather variability needs to be predicted
correctly for the distribution of loss to be predicted. Weather that
is highly correlated in time is more likely to lead to a run of
freezing days and hence a very high loss.
Correlation is not perfectly forecasted by numerical weather prediction
models, and, like all other aspects of forecasts, needs calibration (\citet{jewson03aa}).
The catch is that it is very difficult, if not impossible, to perform
correlation calibration and still preserve the ensemble members.
Thus if one restricts oneself to the end to end use of ensembles one can
never avail oneself of an optimum forecast of the correlation.

\subsubsection{The ensemble size problem}
For some users of weather forecasts the extreme tails of the distribution
of future weather are important. By definition, these are not well
sampled by ensemble forecasts with only a small number of members.
For instance, consider a business that goes bankrupt if it freezes during
the next 10 days, but that freezing has a probability of 1 in 100.
End to end use of ensembles for ensemble sizes that are not much greater
than 100 will not estimate the probability of bankrupty very well.

\section{Full statistical modelling of ensemble forecasts}

The limitations of end to end use of ensembles can be overcome using
what we will call the 'full statistical modelling method'. This method
works by 'deconstructing' the ensemble completely into a probability
distribution (or probabilistic forecast). The original members
are lost, and all that remains is the estimated distribution of future
weather. This distribution is estimated as an optimal combination of
information from the ensemble and from past forecast errors using methods
such as those that we describe in~\citet{jewson04l}.

The advantages of this method are a) the entire distribution of future temperatures
can be predicted in an optimal way, making use of all available information including
correlation information from past forecast error statistics and
b) the ensemble size can be increased as large as is needed (by resampling the forecast
distribution) and can hence capture
low probability events.

\section{Summary}

We have discussed the end to end and full statistical modelling methods for making
use of the information present in weather forecast ensembles. We find that the end
to end method has two shortcomings: sub-optimal correlation forecasts and limited
ensemble size. These are both overcome by using the full statistical modelling method.

\section{Acknowledgements}

Thanks to Christine Ziehmann for some interesting discussions on this topic.

\section{Legal statement}

SJ was employed by RMS at the time that this article was written.

However, neither the research behind this article nor the writing
of this article were in the course of his employment, (where 'in
the course of their employment' is within the meaning of the
Copyright, Designs and Patents Act 1988, Section 11), nor were
they in the course of his normal duties, or in the course of
duties falling outside his normal duties but specifically assigned
to him (where 'in the course of his normal duties' and 'in the
course of duties falling outside his normal duties' are within the
meanings of the Patents Act 1977, Section 39). Furthermore the
article does not contain any proprietary information or trade
secrets of RMS. As a result, the author is the owner of all the
intellectual property rights (including, but not limited to,
copyright, moral rights, design rights and rights to inventions)
associated with and arising from this article. The author reserves
all these rights. No-one may reproduce, store or transmit, in any
form or by any means, any part of this article without the
author's prior written permission. The moral rights of the author
have been asserted.

The contents of this article reflect the author's personal
opinions at the point in time at which this article was submitted
for publication. However, by the very nature of ongoing research,
they do not necessarily reflect the author's current opinions. In
addition, they do not necessarily reflect the opinions of the
author's employer.

\bibliography{end2end}

\end{document}